\documentclass[twoside]{dis04}

\usepackage{axodraw}
\usepackage{amsmath}

\def\runauthor{I.~Schienbein}
\def\shorttitle{$D^\star$ Hadroproduction with Massive Quarks}

\def\Fortran{{\tt FORTRAN\,}}

\def\msbar{\ensuremath{{\rm{\overline{MS}}}}}

\newcommand{\gev    }{\ensuremath{\mathrm{GeV}}}
\newcommand{\tev    }{\ensuremath{\mathrm{TeV}}}

\newcommand{\der}{\ensuremath{{\operatorname{d}}}}
\newcommand{\Ord}{\ensuremath{{O}}}

\newcommand{\cbar}{\ensuremath{\bar{c}}}
\newcommand{\qbar}{\ensuremath{\bar{q}}}

\newcommand{\pbar}{\ensuremath{\bar{p}}}

\newcommand{\mui}{\ensuremath{\mu_f}}
\newcommand{\muf}{\ensuremath{\mu_f^\prime}}
\newcommand{\mur}{\ensuremath{\mu_r}}

\newcommand{\sub}{\ensuremath{\rm SUB}}

\begin{document}

\title{INCLUSIVE $D^\star$ HADROPRODUCTION WITH MASSIVE QUARKS}

\author{I.~SCHIENBEIN}

\address{DESY, \\
Notkestrasse 85, \\ 
22603 Hamburg, Germany\\ 
E-mail: schien@mail.desy.de}

\maketitle

\abstracts{
A calculation of the next-to-leading order cross section for the
inclusive hadroproduction of charm quarks including heavy quark mass 
effects in the hard scattering cross sections is presented.
It is described how the massive hard scattering cross sections 
are constructed from the corresponding cross sections in a 
fixed order calculation where collinear pieces associated with
the heavy quark are not yet subtracted.
By adjusting suitable subtraction terms
a massive theory with $\msbar$ subtraction is established which
approaches the massless theory with increasing transverse
momentum. 
These results will allow to have a more solid comparison 
with recent data for the inclusive $D^{\star\pm}$ cross section
in $p\pbar$ collisions from the CDF collaboration at the Tevatron 
at $\sqrt{S} = 1.96\ \tev$. 
}

\section{Introduction}
\label{sec:intro}
Heavy quarks are those with masses
$m_h \gg \Lambda_{\rm QCD}$ such that $\alpha_s(m_h^2) \propto
\ln^{-1}(\tfrac{m_h^2}{\Lambda^2_{\rm QCD}}) \ll 1$ and 
hence, according to this definition, the charm, bottom and top quarks 
($c, b, t$)
are heavy whereas the up, down and strange quarks ($u, d, s$) 
are light.
Apart from the fact that heavy quark production
processes are fundamental elementary
particle processes which take place with substantial rates
at high energy colliders, 
and which therefore require a good phenomenological understanding,
heavy quark production is theoretically interesting for 
at least two more reasons.
(i) Firstly, the heavy quark mass $m_h$ acts as a physical 
long-distance cut-off such that the heavy quark (or more precisely
logarithms of the heavy quark mass) can be treated
within perturbative QCD -- either within fixed order perturbation
theory or in resummed approaches where a heavy quark parton distribution
function (PDF) is introduced using {\em perturbatively calculated}
boundary conditions for the Altarelli-Parisi evolution equations.
%
(ii) Secondly, in addition to $m_h$ another
large scale, e.g. the transverse momentum $p_T$ of the heavy quark, 
is usually involved in the hard scattering process such that
one has to deal with multi-scale processes which complicate the
perturbative analysis. 
Obviously, terms of the order $\Ord(m_h^2/p_T^2)$ should not be neglected
in resummed approaches if $p_T$ is not much larger than the 
heavy quark mass.
This remains also true if the perturbative boundary conditions in (i) 
would be abandoned in favour of experimentally determined 
input distributions for the heavy quark PDF.

The fixed order treatment is also called fixed flavour number
scheme (FFNS) since the number of flavours in the initial state
is fixed to $n_f=3(4)$ for charm (bottom) production.
On the other hand, in the parton model the number of active flavours
is increased by one unit, $n_f \to n_f +1$, if the factorization scale
crosses certain transition scales (which are usually taken to be 
of the order of the heavy quark mass)\footnote{For a detailed 
discussion see the appendix 
in \cite{Amundson:2000vg} and references given there.}. 
Accordingly, these schemes are called
variable flavour number schemes (VFNS).
The conventional massless parton model is usually referred to as
massless or zero mass VFNS (ZM-VFNS)  whereas massive parton 
model approaches are named massive or general mass VFNS (GM-VFNS).
For details see, e.g., Refs.\ \cite{Tung:2001mv,Schienbein:2003et}.

In this contribution I will consider open charm 
hadroproduction, $p \pbar \to D^\star X$, 
and describe a new calculation 
in a massive VFNS \cite{Kniehl:winp} 
guided by the
factorization theorem of John Collins including 
heavy quark masses \cite{Collins:1998rz}.
This work is along the lines of previous studies
of the direct \cite{Kramer:2001gd} and 
the single-resolved contributions \cite{Kramer:2003cw} 
to the process $\gamma \gamma \to D^\star X$.
Note also that the latter constitutes the direct contribution to
$\gamma p \to D^\star X$ considered in 
\cite{Kramer:2003jw,Schienbein:2003et}.
The results of this new calculation can then be used also to obtain the
double resolved contribution in the $\gamma\gamma$ and the resolved
contribution in the $\gamma p$ process in order to perform
a complete NLO analysis of these processes in this massive VFNS.

Recently, the CDF collaboration has published first cross section data
for single inclusive $D$ meson production 
in $p\pbar$ collisions \cite{Acosta:2003ax} obtained in Run II at
the Tevatron at center-of-mass energies of $\sqrt{S} = 1.96\ \tev$.
These data have been compared with calculations \cite{Cacciari:2003zu,BAK} 
in two distinct theoretical approaches, the FONLL approach
\cite{Cacciari:1998it} and the conventional massless parton model
\cite{BKK-D}, which both are compatible with the data
within theoretical and experimental errors.\footnote{However,
for the central values the ratio of data/theory is about $2$ at low 
$p_T$ and $1.5$ at high $p_T$.}
The work presented here \cite{Kniehl:winp} extends the 
latter successful calculation in the massless approach by 
terms of the order $\Ord(m_c^2/p_T^2)$.

\section{Basic Framework}
\label{sec:framework}

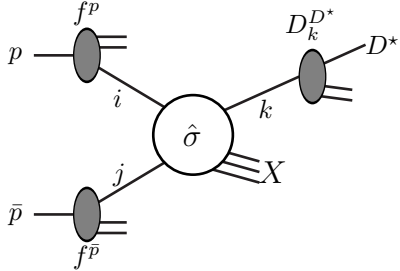
\begin{figure}[t]
\begin{center}
{\begin{minipage}[]{0.41\linewidth}
\begin{picture}(140,100)(-10,-10)
\SetScale{1.0}
\SetWidth{1.0}
\Line(0,10)(15,10)
\Line(20,7)(35,7)
\Line(20,3)(35,3)
\Line(20,12)(50,30)
{\GOval(20,10)(10,5)(0){0.5}}

\Line(0,70)(15,70)
\Line(20,73)(35,73)
\Line(20,77)(35,77)
\Line(20,68)(50,50)
{\GOval(20,70)(10,5)(0){0.5}}

\Line(60,37)(85,30)
\Line(60,33)(85,26)
\Line(60,29)(85,22)
\Line(60,44)(100,62)
{\GOval(60,40)(15,15)(0){1.0}}

\Line(105,59)(120,57)
\Line(105,55)(120,53)
\Line(105,65)(125,74)
{\GOval(105,62)(10,5)(0){0.5}}

\Text(30,25)[l]{\normalsize $j$}
\Text(30,54)[l]{\normalsize $i$}
\Text(85,50)[l]{\normalsize $k$}
\Text(85,26)[l]{\large $X$}

\Text(60,40)[c]{\large $\hat{\sigma}$}

\Text(20,-6)[c]{\normalsize $f^{\pbar}$}
\Text(20,86)[c]{\normalsize $f^p$}
\Text(105,82)[c]{\normalsize $D_k^{D^\star}$}

\Text(132,74)[c]{\normalsize $D^\star$}
\Text(-7,10)[c]{\normalsize $\pbar$}
\Text(-7,70)[c]{\normalsize $p$}
\end{picture}
\end{minipage}} 
\end{center}
\caption{Open charm hadroproduction, $p \pbar \to D^\star X$, according
to Eq.\ \eqref{eq:hadfact}.}
\label{fig:hadfact}
\end{figure}

Our theoretical basis for the calculation of differential 
cross sections for open charm 
hadroproduction, $p \pbar \to D^\star X$, 
is the familiar factorization formula [see Fig.~\ref{fig:hadfact}],
however, with heavy quark mass terms included in the hard scattering
cross sections \cite{Collins:1998rz}:
\begin{eqnarray}
\der \sigma(p \pbar \to D^\star X) 
& =& 
\sum_{i,j,k} 
f_i^p(x_1,\mui)\otimes f_j^{\pbar}(x_2,\mui)\otimes 
\der \hat{\sigma}(i j \to k X)\otimes D_k^{D^\star}(z,\muf)\, .
\label{eq:hadfact}
\end{eqnarray}
Here $f_i^p(x_1,\mui)$ and $f_j^{\pbar}(x_2,\mui)$,
$i,j=u,d,s,c,g$, are universal PDFs of the proton
and the anti-proton 
which are non-perturbative quantities in the case of gluons, $i,j = g$,
and light quarks, $i,j=q \equiv u,d,s$.
On the other hand, the charm quark PDF is usually obtained from
perturbatively calculable boundary conditions.
Furthermore, the fragmentation of the final state parton 
$k=g, q, c, \ldots$ into the
$D^\star$ meson is described by universal 
non-perturbative fragmentation functions (FFs) $D_k^{D^\star}(z,\muf)$.
Finally, the hard scattering cross sections
$\der \hat{\sigma}(i j \to k X)(\mui,\muf,\alpha_s(\mu),\tfrac{m_c}{p_T})$ 
are perturbatively calculable and depend on the factorization
scales $\mui$ and $\muf$ and the renormalization scale $\mu$.
As has been shown in Ref.\ \cite{Collins:1998rz}, heavy quark
mass effects can be consistently included in the hard part
where it is of course mandatory to use the same factorization
scheme for PDFs, FFs and the hard part.

\section{Hard scattering coefficients with masses}
\label{sec:hsc}

For the calculation of massive hard scattering coefficients
we have adopted the following procedure 
\cite{Kramer:2001gd,Kramer:2003cw,Schienbein:2003et}:
(i)
We have calculated the $m \to 0$ limit of the massive 3-FFNS 
coefficients \cite{Bojak:2001fx,Bojak-PhD}
only keeping $m$ as regulator in logarithms $\ln \tfrac{m^2}{s}$.
The partonic subprocesses occurring in the 3-FFNS are
$g g \to c \cbar$ and $q \qbar \to c \cbar$ in LO and
$g g \to c \cbar g$, $q \qbar \to c \cbar g$, and
$g q \to c \cbar q$ ($g \qbar \to c \cbar \qbar$) in NLO.
Here special care was required in order to recover certain distributions
($\delta(1-w)$, $(\tfrac{1}{1-w})_+$, $(\tfrac{\ln(1-w)}{1-w})_+$)
occurring in the massless $\msbar$ calculation. 
%
(ii)
Subsequently, we have compared the massless limit with the
corresponding short distance coefficients in the
genuine massless $\msbar$ calculation 
in order to identify the subtraction terms by the unique prescription
\begin{equation}
\der \sigma_{\sub} = ( \lim_{m \to 0} \der \sigma(m) )
- \der \hat{\sigma}(\msbar)\ .
\end{equation}
The massive hard scattering cross sections have then been obtained
by removing the subtraction term from the massive coefficient in
the 3-FFNS:
\begin{equation}
\der \hat{\sigma}(m) = \der \sigma(m) - \der \sigma_{\sub} \, .
\end{equation}
It should be noted that
by this procedure no finite mass terms are removed from the
hard part apart from the collinear logarithms $\ln \tfrac{m^2}{s}$.
%
(iii)
Contributions with charm quarks in the 
{\em initial state} have been included in the massless 
approach \cite{Kramer:2000hn}.
Note that this can be done without loss of precision \cite{Kramer:2000hn}.
Moreover, this rule is of great practical importance 
since the existing massless 
results for the hard scattering coefficients \cite{Aversa:1988vb} 
can simply be used whereas their massive analogues are unknown
and would require a dedicated calculation of these processes.
Note that massive heavy quark initiated coefficients
have been obtained for the case of deep inelastic scattering 
\cite{Kretzer:1998ju,Kretzer:1998nt} and the results are already 
quite involved.
Finally, this rule (scheme) has also been employed
in a recent analysis of CTEQ PDFs with heavy quark 
mass effects \cite{Kretzer:2003it} such that it is
necessary to do the same if these PDFs are to be used
in a consistent analysis in the future.

\section{Numerical Results}
\label{sec:results}

In this section numerical results for the $g g \to c+X$ 
channel (i.e.\ for $p \pbar \to (gg \to c) \to D^{\star}+X$)
are presented
in form of the ratio 
\begin{equation}
 R_{gg}= 
\frac{\int_{-1}^1 \der y\ \der \sigma^{gg}/\der y \der p_T}{\int_{-1}^1 
\der y\ \der \sigma^{gg}_{\rm LO}/\der y \der p_T (m = 0)}\, ,
\label{tratio}
\end{equation}
in dependence of the $p_T$ of the $D^{\star+}$ meson.
In order to investigate the size of the mass effects the cross section
$\der \sigma/\der p_T \der y$ in the numerator has been calculated 
in two ways: 
firstly, in the massive 3-FFNS \cite{Bojak:2001fx}
using the $\Fortran$ code provided by I.~Bojak 
and secondly, using the massless limit which we derived analytically 
as described in the preceding section.
In both cases the LO cross section in the denominator has been
calculated with massless charm quarks.

%
%
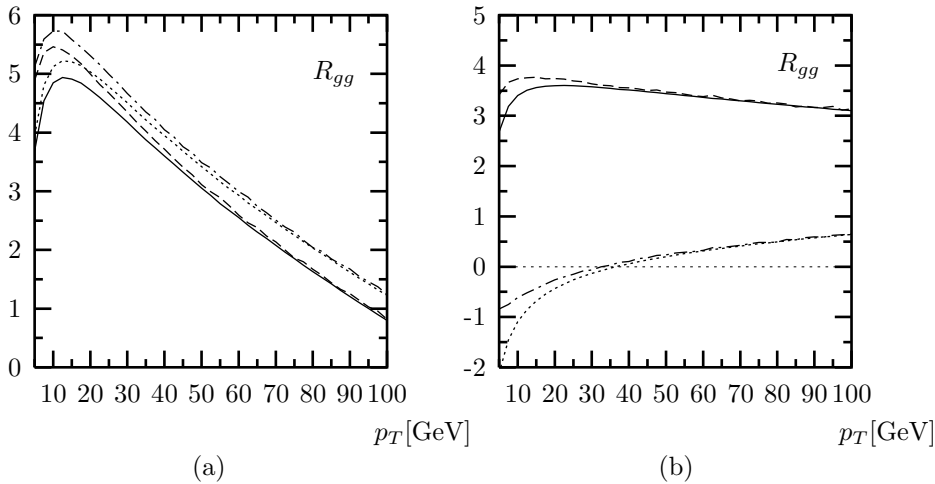
\begin{figure}[ht!] 
\unitlength 1mm
\begin{picture}(140,70)
\put(0,0){\begin{minipage}[b][70mm][b]{70mm}
\label{fig:fig1a}
\include{fig1}
\end{minipage}}
\put(60,0){\begin{minipage}[b][70mm][b]{70mm}
\label{fig:fig1c}
\include{fig3}
\end{minipage}}
\put(28,2){(a)}
\put(90,2){(b)}
\end{picture}
\caption{Contribution of the $g g \to c X$ channel to 
$p\pbar \rightarrow D^{\ast} + X$ normalized to the LO cross
section with $m=0$. 
Shown are results for the massless limit (solid line), the massive
calculation (dashed line), the subtracted massless limit (dotted line),
and the subtracted massive calculation (dashed-dotted line).
The Factorization and renormalization scales are
(a) $\mui = \muf = \mur = m$ (but fixed at $\mu = 2.1m$ in
PDFs, FFs, and $\alpha_s$) and
(b) $\mui = \muf = \mur = 2 m_T$.}
\label{fig:fig1}
\end{figure}

The results in Fig.~\ref{fig:fig1} have been obtained using 
the CTEQ6M PDFs \cite{Pumplin:2002vw} and the FFs for 
$c \rightarrow D^{*+}$ from \cite{Binnewies:1998xq} (NLO OPAL version) 
along with the following input parameters,
$m= 1.5\ \gev$, $E_p=\sqrt{s}/2 = 980\ \gev$, 
$\Lambda_{(n_f = 4)} = 328$ MeV (i.e., $\alpha_s(m_Z) = 0.118$). 
In Fig.\ \ref{fig:fig1} (a),
the renormalization 
and the factorization scales are $\mur=\mui=\muf=m$, but the scale in 
$\alpha_s$, in the PDFs and the FFs is chosen as $\mu=2.1m$, to stay 
above the starting scale in the PDFs and FFs. 
The solid lines 
are the massless limit 
and the dashed curves are the result of the massive calculation.
As one can see, the massive cross section approaches
the massless one very slowly at large $p_T$. 
At $p_T=20\ \gev$ the
difference between the massive and the massless result 
is still of the order of $6\%$ and
the massive cross section is always larger than the massless  
cross section ratio in the $p_T$ range between $7$ and $100\ \gev$. 
From this 
comparison we conclude that in the $gg$ channel the terms proportional to 
$m^2/p_T^2$ are particularly large and lead to an increase of the massive
cross section as compared to the massless approximation. 
This agrees with findings in \cite{Cacciari:1998it}, where the massive 
and the massless theory were compared as a function of the mass $m$ 
for fixed $p_T$. 
For our application we
are interested in the massive cross section and its massless limit, 
where the subtraction terms $\der \sigma^{\sub}$ are removed.
This leads to the dashed-dotted (massive) and the dotted (massless) 
curves in Fig.~\ref{fig:fig1} (a).
Not shown is the genuine $\msbar$ result,
generated with the routine of 
\cite{Aversa:1988vb,Kniehl:1996we,Binnewies:1997gz},
which is in exact agreement
with the subtracted massless limit (dotted curve) 
demonstrating the correctness of the subtraction terms for 
$\mur = \mui = \muf = m$.
%
%
Fig.~\ref{fig:fig1} (b) shows results for the choice
$\mur=\mui=\muf= 2m_T=2\sqrt{m^2+p_T^2}$ which can be used for small 
and large $p_T$. 
The cross sections for these scales are shown  
again in four curves, massless theory, massive 3-FFNS, subtracted 
massless theory and subtracted massive theory where the
labeling of the curves is the same as in Fig.~\ref{fig:fig1} (a). 
These curves are our final results for the $gg$ channel. 
As can be seen,
for this different scale the cross section ratios are much flatter 
as a function of $p_T$ than in Fig.~\ref{fig:fig1} (a). 
The K factor is somewhat smaller now, but it is still 
large, showing that for the $gg$ channel the perturbative expansion is not 
converging very well. 
As in Fig.~\ref{fig:fig1} (a)
we observe that the massive cross section converges from above
to the corresponding massless cross sections with increasing $p_T$.
The ratios for the subtracted cross sections are now
negative for $p_T \leq 30\ \gev$ and rise up to $\approx 0.6$
at $p_T = 100\ \gev$. 
Also for these scales we find exact agreement between
the genuine $\msbar$ result
and the subtracted massless limit (dotted curve).
Thus we can be sure that we have
subtracted from the massive cross section the correct terms in order 
to get the massive VFNS cross section in the $\msbar$ subtraction 
scheme, which approaches the massless limit in this scheme at large $p_T$.

In contrast to the $gg$ channel, the mass effects in the $q\qbar$
and $gq$ channels are quite small reducing the size of the 
mass effects in the sum of all three channels.
Furthermore, contributions with charm quarks in the initial state
and contributions from the gluon (and light quarks) fragmenting
into the $D^\star$ meson have to be included.
These contributions are dominant\footnote{This can be understood 
already with help of Fig.\ \ref{fig:fig1} (b) where large subtractions
are visible which are resummed in the evolved 
charm quark PDFs (and FFs).}
and calculated with $m=0$ as explained
in Sec.\ \ref{sec:hsc}.
This further dilutes the mass effects occurring in the $gg$ channel 
such that quite small and positive mass effects will survive in the 
total result \cite{Kniehl:winp}.


\section*{Acknowledgements} 
I am grateful to B.~A.~Kniehl, G.~Kramer and H.~Spiesberger
for their collaboration and to I.~Bojak for providing his
$\Fortran$ code for heavy quark production in hadron hadron 
collisions. 




\end{document}